\documentclass[conference]{IEEEtran}
\IEEEoverridecommandlockouts
\usepackage{cite}
\usepackage{amsmath,amssymb,amsfonts}
\usepackage{algorithmic}
\usepackage{graphicx}
\usepackage{textcomp}
\usepackage{xcolor}
\usepackage{multirow}
\usepackage{booktabs}
\usepackage{graphicx}
\usepackage{makecell}
\usepackage[printonlyused]{acronym}
\usepackage{listings}
\usepackage{caption} 
\usepackage{threeparttable}

\usepackage[T1]{fontenc}    %
\usepackage{textcomp}      %

\usepackage{hyperref}

\usepackage{tikz}

\newcommand\copyrighttext{%
  \footnotesize \textcopyright 2026 IEEE. Personal use of this material is permitted.
  Permission from IEEE must be obtained for all other uses, in any current or future
  media, including reprinting/republishing this material for advertising or promotional
  purposes, creating new collective works, for resale or redistribution to servers or
  lists, or reuse of any copyrighted component of this work in other works.}
\newcommand\copyrightnotice{%
\begin{tikzpicture}[remember picture,overlay]
\node[anchor=south,yshift=10pt] at (current page.south) 
  {\fbox{\parbox{\dimexpr\textwidth-\fboxsep-\fboxrule\relax}{\copyrighttext}}};
\end{tikzpicture}%
}
\acrodef{AP}{Access Point} %
\acrodef{AOSP}{Android Open Source Project}
\acrodef{API}{Application Programming Interface}
\acrodef{BLE}{Bluetooth Low Energy}
\acrodef{CSI}{Channel State Information}
\acrodef{CV}{Computer Vision}
\acrodef{dBm}{decibel-milliwatt}
\acrodef{GHz}{gigahertz}
\acrodef{HT-LTF}{High Throughput Long Training Field}
\acrodef{IoT}{Internet of Things}
\acrodef{LiDAR}{Light Detection and Ranging}
\acrodef{LR-WPAN}{Low-Rate Wireless Personal Area Network}
\acrodef{MAC}{Medium Access Control}
\acrodef{OPERA}{Object Perception \& Application}
\acrodef{OS}{Operating System}
\acrodef{PHY}{Physical}
\acrodef{PLCP}{physical layer convergence protocol}
\acrodef{PPDU}{physical layer protocol data unit}
\acrodef{R-CNN}{Region-based Convolutional Neural Network}
\acrodef{RF}{radio frequency}
\acrodef{RGB}{Red Green Blue}
\acrodef{RSSI}{Received Signal Strength Indicator}
\acrodef{SDK}{Software Development Kit}
\acrodef{SDR}{Software Defined Radio}
\acrodef{WLAN}{Wireless Local Area Network}
\acrodef{WPA}{Wi-Fi Protected Access}

\begin{document}

\title{Cross-Domain Inference for Human Localization: Applying Wi-Fi RSSI Data to CSI-Trained Models
}

\author{
\IEEEauthorblockN{Ariel Duschanek-Myers, Thomas Welsh, and Helmut Neukirchen}
\IEEEauthorblockA{Department of Computer Science, University of Iceland, Reykjavík, Iceland\\
\{aad10, tomwelsh, helmut\}@hi.is}
}

\maketitle
\copyrightnotice
\begin{abstract}
Wi-Fi signal data can be used to compromise the privacy of individuals. While many existing approaches rely on \ac{CSI}, collecting this data on typical \acs{IoT} devices often requires elevated operating system permissions and specialized drivers. Consequently, this paper investigates the feasibility of utilizing \ac{RSSI} data to predict human locations. \ac{RSSI} was selected because it is accessible even on devices with limited user permissions, and therefore is more applicable to a wider array of \acs{IoT} devices. To bypass the tedious process of obtaining training data needed to train an \ac{RSSI}-based model, an existing Wi-Fi pose prediction project was used in this research. However, that project assumed \ac{CSI} data as input. Therefore, we investigate the feasibility of cross-domain inference, i.e.,\ feeding \ac{RSSI} data into that existing \ac{CSI}-based model.
We collected an \ac{RSSI} dataset, synchronized with video ground-truth of a person moving within a room, to evaluate the model's performance. This evaluation confirmed that \ac{RSSI} data 
can predict locations with approximately 80\% confidence when human movement is present.
This demonstrates that a model trained on \ac{CSI} data can be used to evaluate low-granularity \ac{RSSI} data consisting of \ac{dBm} values to roughly locate people in the collection space. These results imply that a wide range of IoT devices can be used for privacy invasion in Wi-Fi-dense environments.
\end{abstract}

\begin{IEEEkeywords}
IoT,  cross-domain inference, location privacy, Received Signal Strength Indicator (RSSI)
\end{IEEEkeywords}

\acresetall

\section{Introduction}
The internet has changed the definition of what it means to be connected in the modern world, and everyday smart devices have only further reshaped what `connected' means. The wireless communications that the \ac{IoT} ecosystem relies on are often forgotten when evaluating what information is exposed by an \ac{IoT} device. Recent research has identified a wide range of personal data incidentally exposed by these devices, such as location or vital signs~\cite{11421930}. The popular adoption of \ac{IoT} devices substantially increased the number of interactions between \ac{RF} signals and physical objects, including people. Several works have shown that fluctuations in \ac{RF} signal quality can be used to identify objects and their movements, with more advanced approaches able to identify human pose, accurate location, and even heartbeat~\cite{9008282,10656837,RobustDeviceFree_ResearchGate,ThroughWallSensing_IEEE,heartbeatwifi}. 

These works indicate that collecting and analyzing \ac{RF} has strong implications for privacy invasion, particularly within the context of an increasing number of deployed \ac{IoT} devices. However, the practical implications of these approaches are unclear due to testing and development under laboratory conditions. For example, these approaches may require access to data (e.g.\ \ac{CSI}~\cite{10737138}) and model training and inference capabilities unobtainable by resource-constrained and commodity IoT devices. 

The contribution of this paper is two-fold: first, investigating the feasibility of using basic Wi-Fi signal data, in the form of \ac{RSSI}, for human localization. Such data can be more easily collected via a smart device's \ac{SDK} than more complex data such as \ac{CSI} which requires lower-level hardware access permissions, typically inaccessible to developers. An example would be Android-based smart TVs that are ubiquitous and could therefore be used to surveil people. The second contribution of this paper is that to understand if \ac{RSSI} data can be used for this purpose, we made use of a \ac{CSI}-trained model for human pose detection~\cite{10656837} with \ac{RSSI} data collected in a similar laboratory setup. We evaluated this \textit{cross-domain inference} \cite{9134370} approach and determine that \ac{RSSI} data can be used to detect human presence and movement in this context. 

The rest of this paper is structured as follows. Section~\ref{sec:RelatedWork} covers background and related work, while Section~\ref{sec:Crossdomain} details our cross-domain inference methodology and experimental setup. Section~\ref{sec:results} and~\ref{sec:Discussion} present results and discuss their implications, respectively. Finally, Section~\ref{sec:Conclusion} concludes this paper.

\section{Background and Related Work}
\label{sec:RelatedWork}

In wireless sensing, the \acf{RSSI} provides a coarse, easily accessible scalar value representing overall signal power at the \ac{MAC} layer. While \ac{RSSI} reflects the aggregate power of the entire channel, \ac{CSI} provides a granular breakdown of that signal. \acf{CSI} is a physical-layer metric that captures amplitude and phase shifts across multiple subcarriers~\cite{11421930}. While \ac{CSI}'s high resolution makes it ideal for advanced perception tasks, it requires specialized network hardware and elevated operating system permissions. In contrast, \ac{RSSI} can be easily obtained by an application program.

The work by Wang et al.~\cite{9008282} was the first that inspired the continued investigation into the use of Wi-Fi for locating people in a space. It focuses on using \ac{CSI} data collected between three pairs of  transmitting and receiving antennas in conjunction with the combined ground truths of a Mask \ac{R-CNN} and OpenPose~\cite{Cao2019OpenPose} models to validate the predictions made by their novel  \emph{Person-in-WiFi} model. 
Their research acknowledges the existing use of Wi-Fi-based perception for persons in a space, but focuses on making this perception more fine-grained with the application of \ac{CSI} data and \ac{CV} models.

Yan et al.~\cite{10656837} continued work on \emph{Person-in-WiFi} with the goal of performing perception in three dimensions. By increasing the number of receivers, the three-dimensional nature of Wi-Fi's broadcast capabilities can be taken advantage of, permitting better estimation where people or limbs overlap. They implemented a new method of generating ground truths by utilizing Microsoft Azure Kinect-generated~\cite{Bertrametal2023} meshes and the Hungarian Loss function to provide three-dimensional representations for the validation of \emph{Person-in-WiFi 3D} predictions. This increases the flexibility of the ground-truth model when it comes to validating the predictions of the \mbox{\emph{Person-in-WiFi 3D}} model by allowing the re-orientation of the mesh in three dimensions as the validation is performed. It continues using \ac{CV} models for both parts of the \emph{teacher-student model} for knowledge transfer while decreasing brittleness.

\begin{figure}[t]
    \centering

    \includegraphics[width=1.0\linewidth]{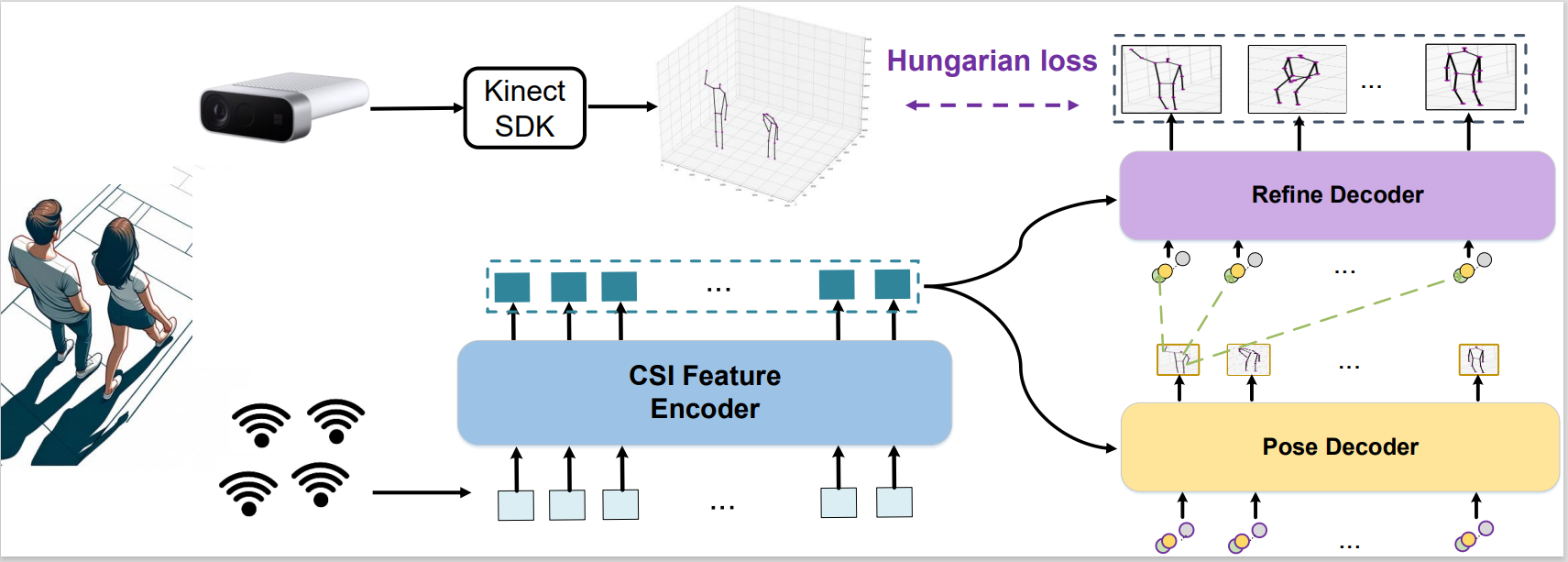}
    \caption{Training environment for the \emph{Person-in-WiFi 3D} model end-to-end \emph{Multi-Person 3D} Pose Estimation with Wi-Fi. (\textcopyright~2024 IEEE. Reprinted, with permission, from Fig.~4 of K.~Yan et al., ``Person-in-WiFi 3D: End-to-End Multi-Person 3D Pose Estimation with Wi-Fi'', 2024 IEEE/CVF Conference on Computer Vision and Pattern Recognition (CVPR)~\cite{10656837}.)}
    \label{fig:Person-in-WiFi_3D_Framework}
\end{figure}

As depicted in \figurename~\ref{fig:Person-in-WiFi_3D_Framework} as the \emph{Pose Decoder}, the student model is trained with \ac{CSI} signal data collected simultaneously as the Kinect captures and labels pose estimations. The pose estimations are used as ground truths during the validation step of the student model's training workflow. The Kinect provides highly reliable pose estimations by using %
depth readings for objects in the space. These estimates allow the student model to tune out background noise and reflected signals.

The \ac{CSI} feature encoder takes the collected data and uses the timestamp to group twenty samples of the 30 channels from each of the 3 antennas for each \ac{AP} together for a frame. A sliding window is used for the frame creation to increase the fluidity of the predicted keypoint movement. The matrix is then passed to a tokenizer that converts the 3 links, 3 antennas, and 20 timepoints into 180 tokens, with each token including a vector of the 30 channel values. This is up-scaled further by 256 by incorporating spatial-temporal embeddings; thus, the model receives a 180 $\times$ 256-dimensional object for evaluation.

Li et al.~\cite{RobustDeviceFree_ResearchGate} provided supporting evidence for the use of \ac{RSSI} data for pose prediction. They focused primarily on the labeling of human poses to provide a minimally invasive method for monitoring elderly occupants of an assisted living space, without the use of video. They performed a comparison between \ac{CSI} and \ac{RSSI} data prediction accuracy for the same number of features (\figurename~8 in~\cite{RobustDeviceFree_ResearchGate}). Though the \ac{CSI} data resulted in roughly a 20 percent increase in accuracy, the \ac{RSSI} data predictions increased in accuracy at the same rate as \ac{CSI} data did when the number of features was increased. This indicates that predictions made based on a high number of features should still provide a reasonably accurate estimate of a generalized position.

Zhu et al.~\cite{ThroughWallSensing_IEEE} utilize a Wi-Fi receiver, a router, and a Received Signal Strength-fit model to detect human movement through a wall. This is done by measuring the Doppler effect that moving objects cause in the broadcast Wi-Fi radio frequency's Received Signal Strength. First, the transmitter is placed in a stationary position in a room and the receiver is moved along a defined path to confirm that the signal strength can be used to locate the transmitter based on changes in the measured signal strength along the path. After the model is fit to the \ac{RSSI} data of the stationary transmitter, the change in signal strength between a stationary transmitter and receiver is correlated with object movement -- in this case, people. The receiver relies on a highly directional antenna and \ac{SDR} to collect \ac{RF} data across the Wi-Fi spectrum more broadly than a standard adapter would permit.

The discussed related work focuses on \ac{CSI} data collected with a specialized implementation in a desktop \ac{OS}, relying on the host's network interface to handle requests for information from the system's Wi-Fi adapter. This is normally not available in consumer \ac{IoT} devices without using specialized tools to gain root access. Though Li et al.~\cite{RobustDeviceFree_ResearchGate} perform a comparison with \ac{RSSI} data, the effect of increased features is documented but not investigated. If an existing solution with a higher number of features is adapted for use with \ac{RSSI} data, it is possible the prediction confidence levels will be competitive with that of models based on \ac{CSI} data.

Besides being \ac{CSI}-based, the related work also focuses on poses as the output of the model's evaluation process. Poses are dependent on the model recognizing smaller volumes of mass than an entire person, something that is useful but resource-intensive. By shifting the focus to location instead of pose, one can rely on the average location of the entire person's mass. This permits the use of sparse data, such as \ac{RSSI} signal strength for evaluation and predictions. Moreover, if the predictions are well correlated and have a small amount of standard deviation, a lower score than is permitted for detecting a pose may still be valid. Therefore the rest of this paper seeks to address these gaps in literature by assessing if RSSI data can be used for cross-domain inference with a CSI trained model to permit human localization.

\section{Cross-domain Inference CSI Model with RSSI Input}
\label{sec:Crossdomain}
To confirm that the \ac{RSSI} data exposed in IoT devices is effective with an existing pose prediction model, this section presents a method for cross-domain inference of a \ac{CSI}-based model using \ac{RSSI} input. The methodology for evaluating this approach is as follows. First, we configure a laboratory aligned with the methodology in~\cite{9008282} and use it to collect \ac{RSSI} signal data affected by human poses and movement. Next, we pre-process the collected data through normalization and noise reduction. Finally, we input this data into the existing \ac{CSI}-based model so that the output can be evaluated. This evaluation considers the model's confidence scores against the ground truth of the human's movements recorded via video.

\subsection{Laboratory Setup}
\begin{figure}[!b]
    \centering\vspace*{-2ex}
    \includegraphics[width=.75\linewidth]{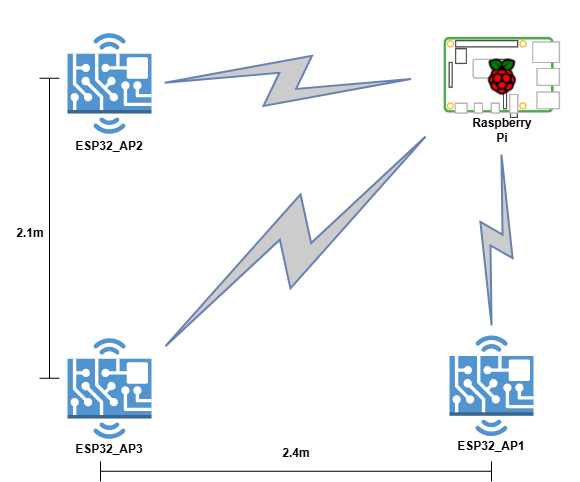}
    \caption{Lab layout}
    \label{fig:lab}
\end{figure}

The \emph{Person-in-WiFi 3D} model~\cite{10656837} that we re-use was trained on a specific lab setup; a roughly 3.5~m by 4~m space with three Wi-Fi routers receiving transmission from one Wi-Fi-enabled laptop. Therefore, we replicated this setup in a campus study space with chairs and tables removed. We placed ESP32-C3 boards acting as \acp{AP} and a Raspberry Pi 4 in the space, along with the necessary power strips (\figurename~\ref{fig:lab}). 

The Raspberry Pi 4 was used in place of the Wi-Fi-enabled laptop. After creating an Android application to investigate whether the required \ac{RSSI} data collection was feasible, the implementation for the receiving device was moved to Linux for simplicity (avoiding lengthy Android app installation procedures) and to ensure empirical consistency. As this work is focusing upon privacy violations in smart environments, the Raspberry Pi was chosen for its similarity to a smart device (such as a smart TV that would, e.g., run Android).

The ESP32-C3 boards serve as simple \ac{AP}s transmitting on a single channel, which is sufficient since signal strength is largely independent of direction in this setup. While this approach may reduce the ability to classify specific joints (i.e.\ pose) of a human body due to the loss of multi-channel data, the positional predictions in the 3D space should be maintained. As depicted in \figurename~\ref{fig:lab}, the access points and the receiving device are within broadcast range of each other; we maintained the scale expected by the model. Three access points are necessary for 3D positional predictions. A single access point provides only one-dimensional information about material obstruction between it and the receiver. With three access points, we can make three-dimensional predictions based on signal strength variations caused by changes in materials between each transmitter and receiver. 

To collect the data, a shell script scans the \ac{RSSI} value on the WiFi interface of the device\footnote{The device operating system was Raspbian 5.6 which is a distribution of Debian Linux with drivers and configuration specific to the Raspberry Pi. The Linux kernel version for this Raspbian build was 6.12.25 and was selected to avoid Wi-Fi driver latency present in later, recent versions.} and stores them  as comma-separated values to be used for model prediction. This script is then implemented as a systemd service which has been configured to run after the network stack is online. The process is then initiated after the device is left stationary in the room.

\begin{figure}[!b]
    \centering\vspace*{-2ex}
    \includegraphics[width=1\linewidth]{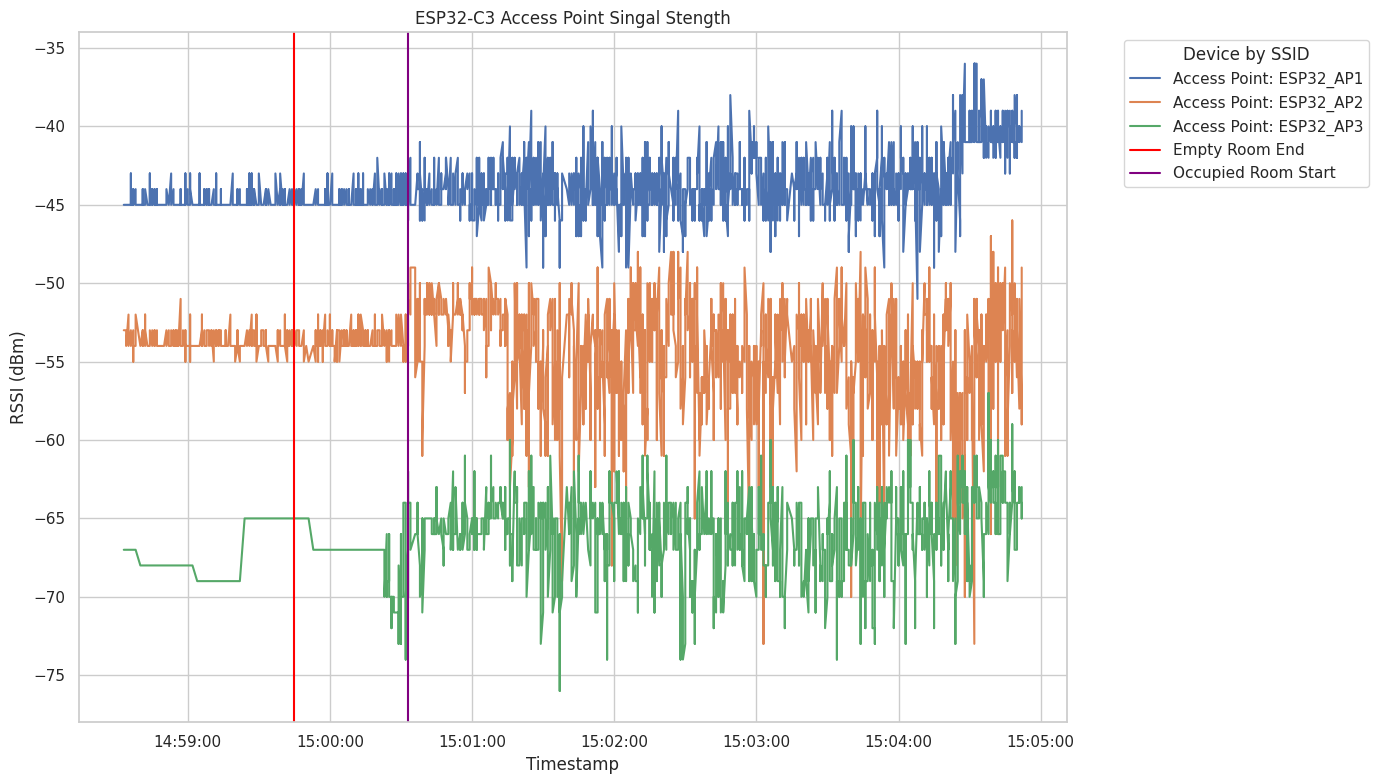}
    \caption{Pre-processing sample RSSI signal strength data}
    \label{fig:sampleData}
\end{figure}

\subsection{Data Collection}

\ac{RSSI} data collection begins in an empty room, with roughly 2 minutes recorded to establish a baseline for correcting environmental noise in the lab. Although the existing \ac{CSI} training data\footnote{\texttt{\url{https://aiotgroup.github.io/Person-in-WiFi-3D/}}} from~\cite{10656837} does not control for interference, the teaching model would reinforce inattentiveness to the background noise during the validation step of training. Ground truths created by the Kinect-labeled meshes are based on infrared and visible light data and therefore cannot include radio interference from the training data collection space, see \figurename~\ref{fig:Person-in-WiFi_3D_Framework}.

A shell script synchronized systems' timestamps via SSH, started the Wi-Fi data collection service on the Raspberry Pi, and began video recording on the laptop. Collection continued until the Enter key was pressed, stopping the Pi service and ending the video recording. This semi-automated workflow allowed us to start collection and leave the room, capturing background noise that was subsequently subtracted to control for changes in the radio frequency environment. 

The baseline \ac{RSSI} data was captured first (empty room with no objects or humans, see the left part of \figurename~\ref{fig:sampleData}, i.e.\ before the vertical Empty Room End line), followed by a second capture period with a person in the space (right part of \figurename~\ref{fig:sampleData}, i.e.\ after the vertical Occupied Room Start line).

\subsection{RSSI Data Preprocessing}

We first focused on signal behavior observed in the empty room. We originally performed testing with two types of \mbox{USB-C} cables and two types of USB wall power adapters. The initial data collection suggested that differences between ESP32\_AP3 and the other two ESP32 \ac{AP}s were caused by the combination of the wall adapter and shielded USB cable. To address this, we collected new data using three identical \mbox{USB-C} cables and wall adapter pairs, such that the noise would become similar for all \ac{AP}s. This required adding surge-protected power strips to power each adapter which standardized the noise from the power sources. The \ac{WPA} Supplicant was disabled to prevent conflicting access to the Wi-Fi adapter by multiple services. A Windows laptop was connected to a USB webcam and the Raspberry Pi via Ethernet cable for remote SSH control. The laptop's Wi-Fi adapter was disabled during collection to prevent possible interference from probe requests. 

We convert the \ac{RSSI} \ac{dBm} signal to float linear values to mimic \ac{CSI} data  before the value was scaled and dampened. This allows us to control for differences in room size of our lab and the lab that was in~\cite{10656837} used for training the model that we re-use, see Listing~\ref{eq:normalization}. 
This linear value is computed as $10^{\frac{\mathrm{RSSI}}{2 \times 10}}$.
To ensure the pre-processing of \ac{RSSI} data collected in our version of the \emph{Person-in-Wifi 3D} lab setup was not artificially improving predictions, we first evaluated the impact of different combinations of normalization (see Table~\ref{tab:confidence_scores}) on the resulting confidence scores that the model provides as part of its output. A second evaluation was performed using only linearized data to verify that normalization has not artificially reduced confidence, since \ac{RSSI} data likely includes some normalization when provided by the Wi-Fi \ac{OS} driver.

\begin{lstlisting}[float=t, language=Python, belowskip=-2ex,
caption={Normalization Python code}, label={eq:normalization}] 
empty_linear = 10 ** (np.array(empty_list) / 20.0)
baseline_amp = np.mean(empty_linear)
if baseline_amp == 0:
    baseline_amp = 1e-9
current_linear = 10 ** (current_rssi / 20.0)
normalized_amp = current_linear - baseline_amp
final_amp = normalized_amp * target_baseline
final_amp = np.power(final_amp, dampen)
return final_amp
\end{lstlisting}

\begin{table}[t]
\centering
  \begin{threeparttable}
\caption{Absolute Maximum Confidence Scores Across Different Data Normalizations. See the footnotes for the normalization applied.}
\label{tab:confidence_scores}
\begin{tabular}{p{1.7cm}cccc}
\toprule
\textbf{Data Type} & \textbf{\makecell{Each AP \\ Baseline\tnote{1}}} & \textbf{\makecell{Each AP \\ Baseline \\ Z-Score\tnote{2}}} & \textbf{\makecell{Tilt \\ Gradient\tnote{3}}} & \textbf{\makecell{ Global Mean \\ Subtraction\tnote{4}}} \\ 
\midrule
Noisy Data\tnote{5} & $0.5774$ & $0.5405$ & $0.7025$ & $0.7922$ \\
Dampen Data\tnote{6} & $0.5492$ & $0.5112$ & $0.7370$ & $0.7904$ \\
Scale Data\tnote{7} & $0.6355$ & $0.3633$ & $0.6252$ & $0.7237$ \\
Norm. Data\tnote{8} & $0.4333$ & $0.3501$ & $0.4997$ & $0.6089$ \\
Empty Data\tnote{9} & $0.3996$ & $0.3978$ & $0.5716$ & $0.6529$ \\
\makecell[l]{Empty Norm. \\Data\footnotemark[10]} & $0.3356$ & $0.3319$ & $0.2750$ & $0.3264$ \\
\bottomrule
\end{tabular}
    \begin{tablenotes}
    \footnotesize
    \item[1] \texttt{current\_linear\_for\_ap / baseline\_amp\_for\_ap}
    \item[2] \texttt{((current\_linear\_for\_ap-g\_mean)/g\_std)+1.0}
    \item[3] \texttt{current\_linear\_for\_ap*(0.9,1.1,num\_subcars)}
    \item[4] \texttt{current\_linear\_for\_ap - baseline\_amp}
    \item[5] \texttt{target\_baseline=1.0, dampen=1.0}
    \item[6] \texttt{target\_baseline=1.0, dampen=0.8}
    \item[7] \texttt{target\_baseline=0.6, dampen=1.0}
    \item[8] \texttt{target\_baseline=0.6, dampen=0.8}
    \item[9] \texttt{target\_baseline=1.0, dampen=1.0}
    \item[10] \texttt{target\_baseline=0.6, dampen=0.8}
    \end{tablenotes}
  \end{threeparttable}
  \vspace*{-2ex}
\end{table}

Predictions were produced by the evaluation pipeline and a comparison was done between the maximum confidence scores, see Table~\ref{tab:confidence_scores}. Based on the difference in the transmitter and room scale, dampening and scaling were applied to the RSSI signal data after linear conversion was performed. Different methods of removing the baseline were tested to determine the best method of removing our specific \ac{RF} background. 

To confirm the need for a filter threshold before data is passed to the model for evaluation, linearized empty room data with and without normalization were evaluated as well. This confirmed noisy spaces with only the median value of the empty room subtracted still provide enough information for the model's attention mechanism to make overly confident predictions, with a maximum confidence of 0.6529. All noise suppression methods besides Per-AP Baseline removal resulted in the empty room data having higher confidence predictions than the normalized, live test data. The visual representation of the data, \figurename ~\ref{fig:sampleData}, makes it apparent that the length and amplitude of changes in signal strength is a useful method for filtering evaluation candidates. 

The maximum confidence score increased from 0.6089 to 0.7922 when the data was left 'noisy' without scale and dampening. If the model were retrained with visual and signal data in the lab space, the reflection and background noise captured in the RSSI signal data would be trained out of relevance and the median and mean confidence of the data with noise would be increased.

A baseline to account for possible changes in the \ac{RF} environment of the lab before the evaluation process was repeated with the \emph{Person-in-WiFi 3D} model, such that the linearized data had the average of the empty room subtracted. We also tested subtracting each \ac{AP}'s average linearized value before applying a noise gate, sigmoid, and angular embedding.

\subsection{Model Integration}

The collected, preprocessed \ac{RSSI} data is fed to the trained model, which produces 100 predictions per frame, each with a confidence score. These scores reflect either successful positive predictions or confident false predictions.

The \emph{Person-in-WiFi 3D} model~\cite{10656837} depends on specific versions of CUDA and PyTorch, which in turn depend on older package versions required by MMDetection, MMCV, and \ac{OPERA}. Upgrading these libraries would necessitate changes to imports and implementation throughout the codebase. To avoid extensive refactoring, we leveraged compatibility options for older versions of Nvidia's CUDA toolkit in combination with the used NVIDIA A100 Tensor Core GPUs.

The output of the evaluation stage of the model is controlled by a configuration parameter, \verb|max_per_img|. We set this to 100 to ensure all 100 query results from the PETRHead stage of the \emph{Person-in-WiFi 3D} model are recorded as output. We also pass the eval\_options `out' parameter a \texttt{.pkl} file name for storing the results.
The result consists of 100 predictions generated for each frame (due to the 100 queries made by the model's evaluation of a frame). A prediction consists of 14 keypoints that represent joints in a humanoid pose and of a score indicating the model's confidence on the accuracy of the prediction of that specific pose.

Since the model outputs 100 different predictions for each analyzed frame, we filter for the highest confidence score to identify the most likely pose for a frame. To validate that this prediction is not an outlier, we plot all 1400 keypoints per frame from the raw output. This visualization reveals whether the prediction confidence score correlates with the model's actual perception of reality as triggered by the evaluated \ac{RSSI} data.

We generate statistics and visualizations for human validation: a graph of prediction score deviation indicates the range of prediction scores. Comparing the maximum score per 100 predictions per frame between the normalized and noisy data will test the theory that \ac{RSSI} data can produce meaningful predictions regardless of the controlled nature of the collection space.  We expected keypoint (=joint) predictions to have low confidence, but the keypoints should be clustered together. %

To determine if the model's confidence scores are accurately predicting real world events, the collected data can be correlated with the video based on the time point recorded with each scan result. Though the scans are meant to be performed twenty times per second, this is an optimistic frequency. As a result of this sampling schedule, frames of pre-processed data are roughly representative of one second. We can therefore plot the prediction frames and compare the video at the likely time code.

\section{Results and Evaluation}\label{sec:results}

\begin{figure}[!b]
    \centering
    \includegraphics[width=0.32\columnwidth]{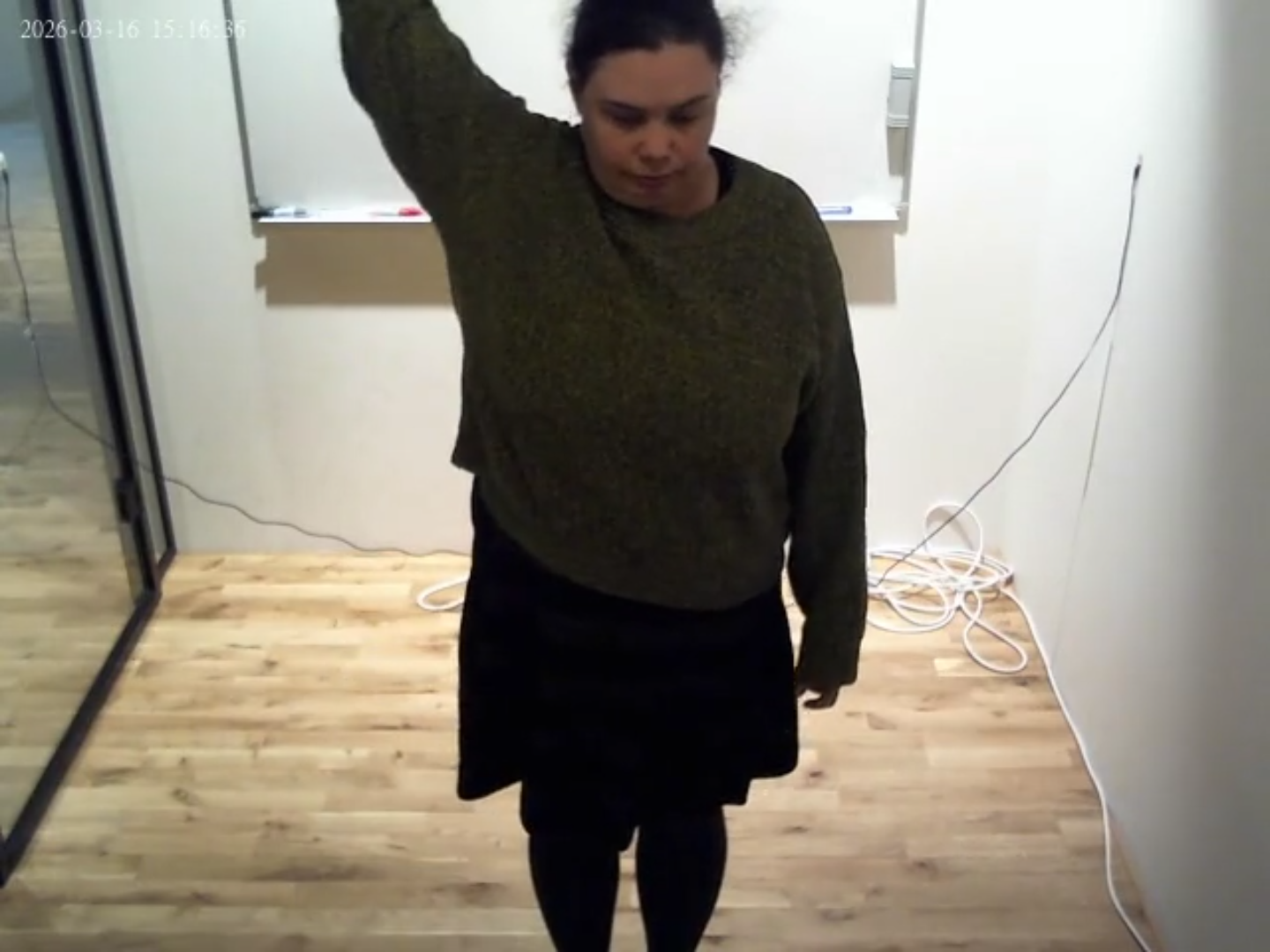}
    \includegraphics[width=0.32\columnwidth]{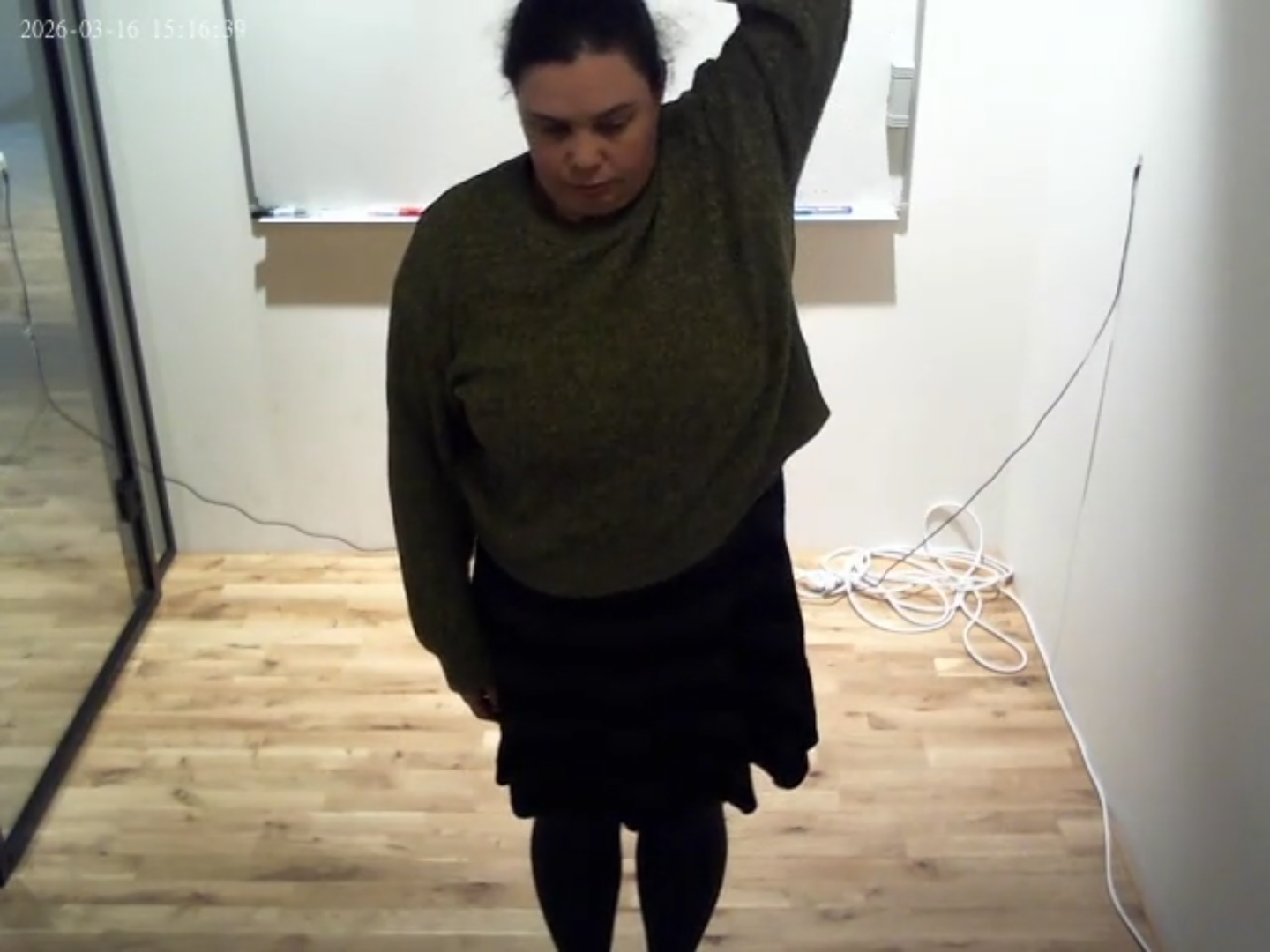}
    \caption*{\small Raising alternating arms}
    \vspace{4mm}
    \includegraphics[width=0.32\columnwidth]{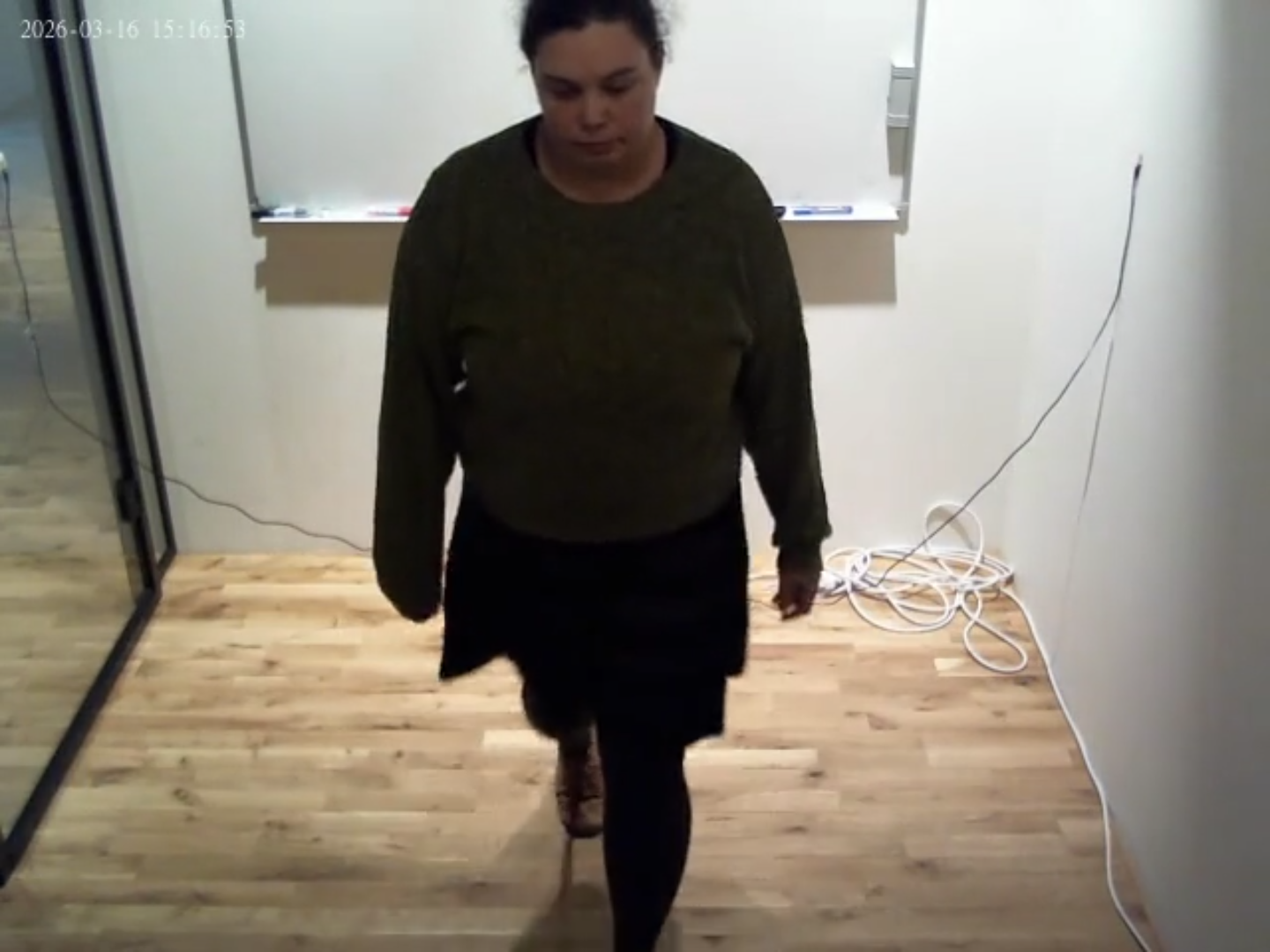}
    \includegraphics[width=0.32\columnwidth]{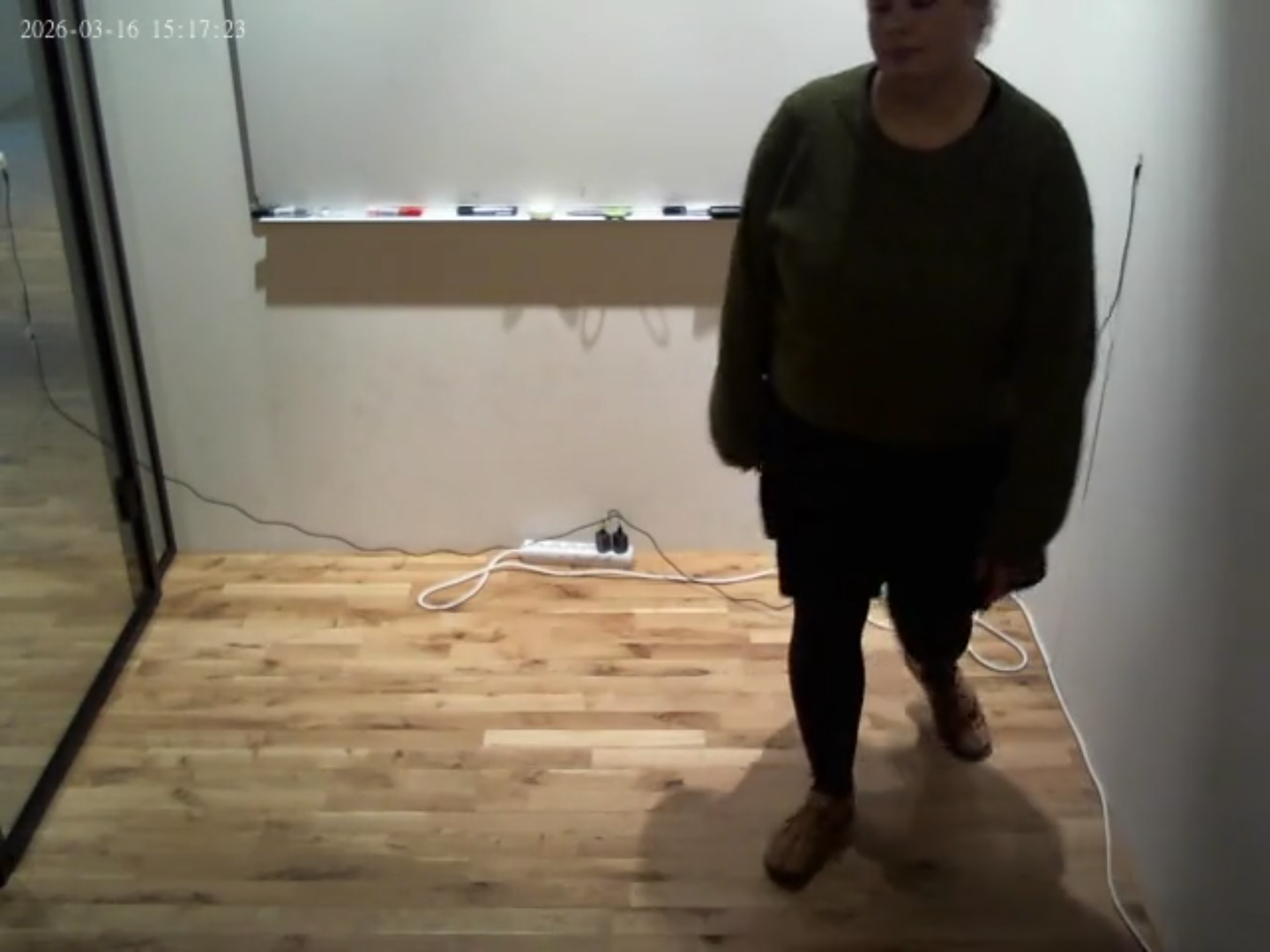}
    \includegraphics[width=0.32\columnwidth]{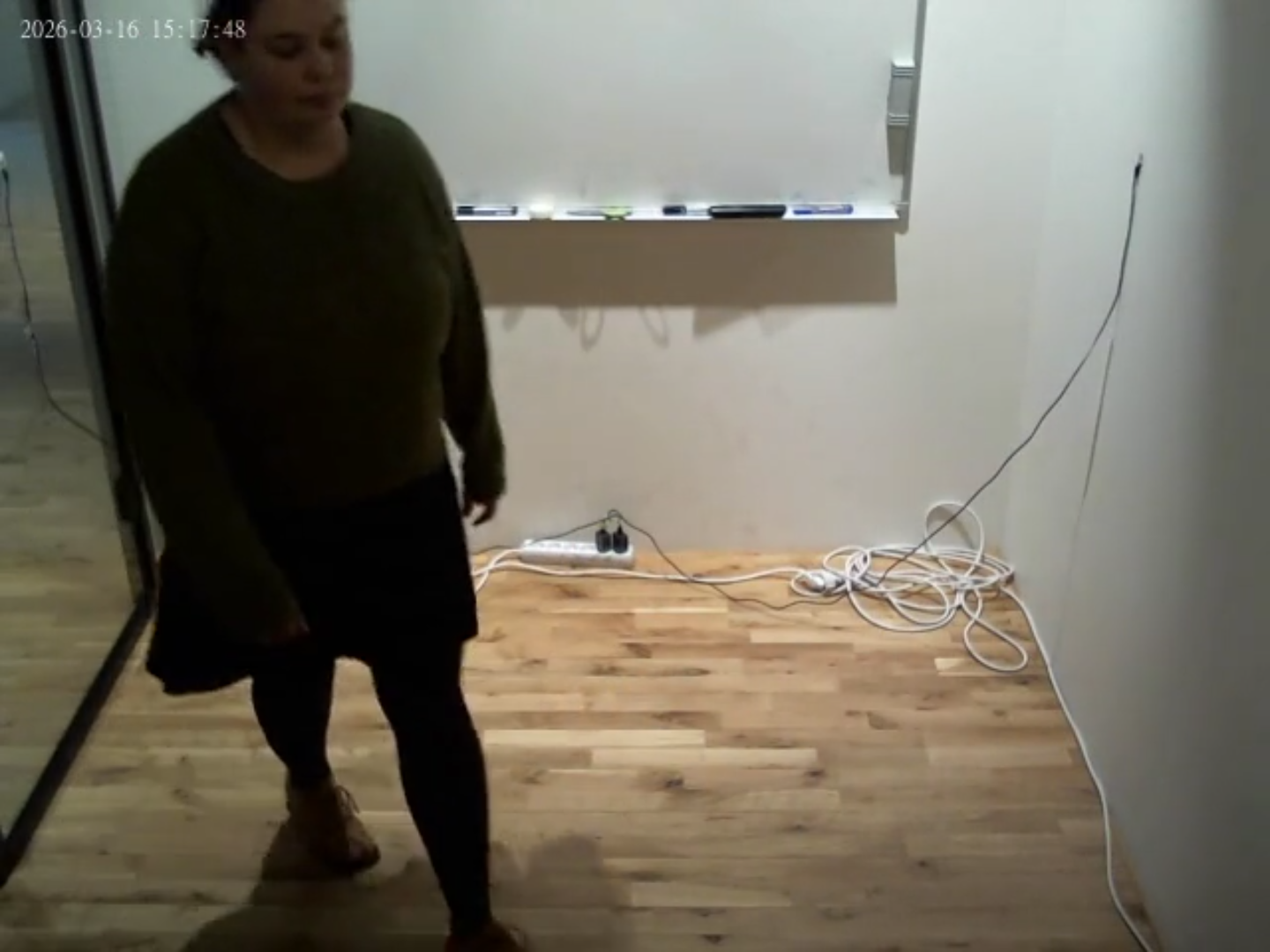}
    \caption*{\small Walking front to back and both diagonals}
    \vspace{4mm}
     \begin{minipage}{0.32\columnwidth}
        \includegraphics[width=\linewidth]{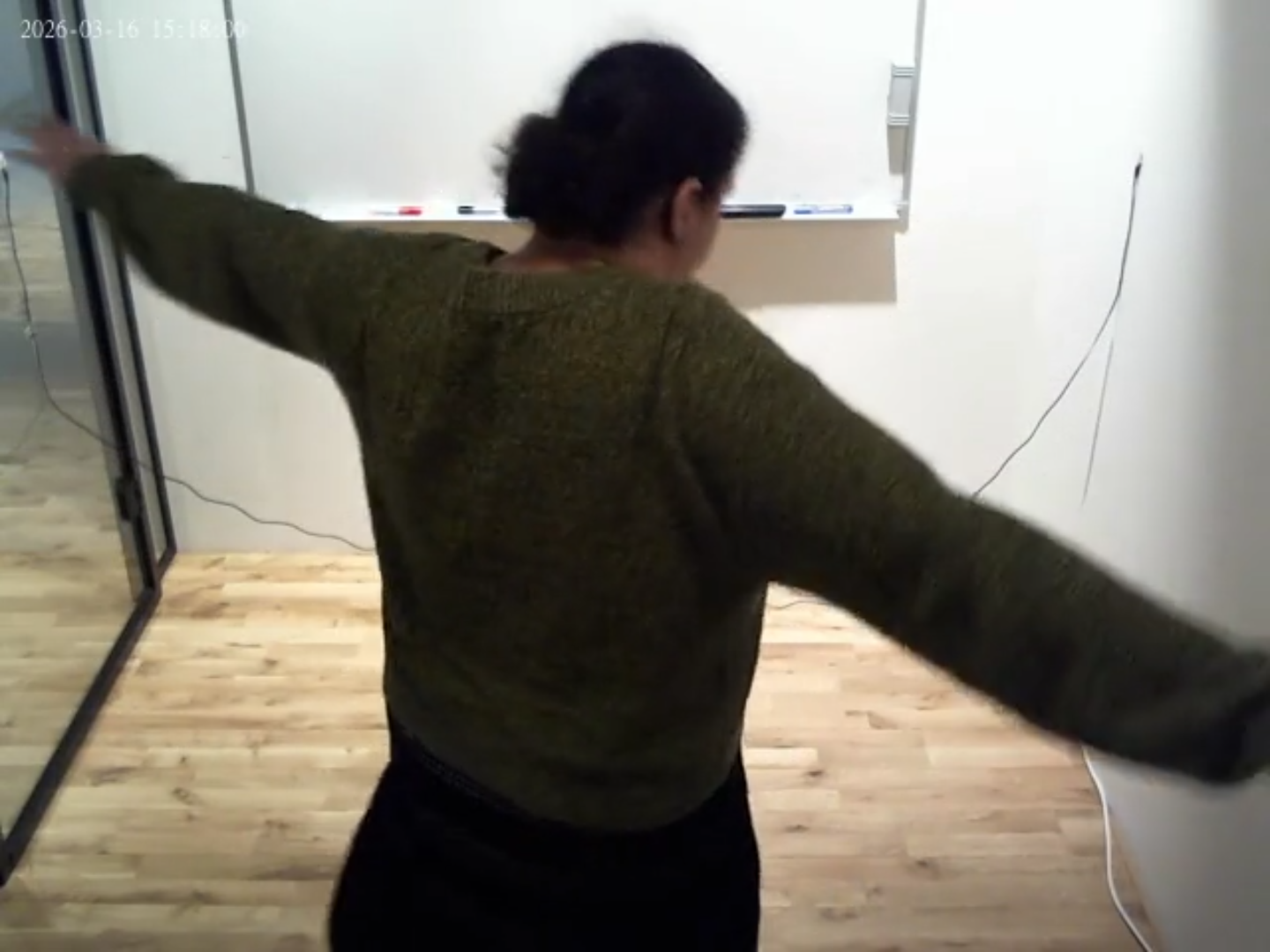}
        \caption*{\small \centering Rotating with \\
        arms straight out}
    \end{minipage}
    \begin{minipage}{0.32\columnwidth}
        \includegraphics[width=\linewidth]{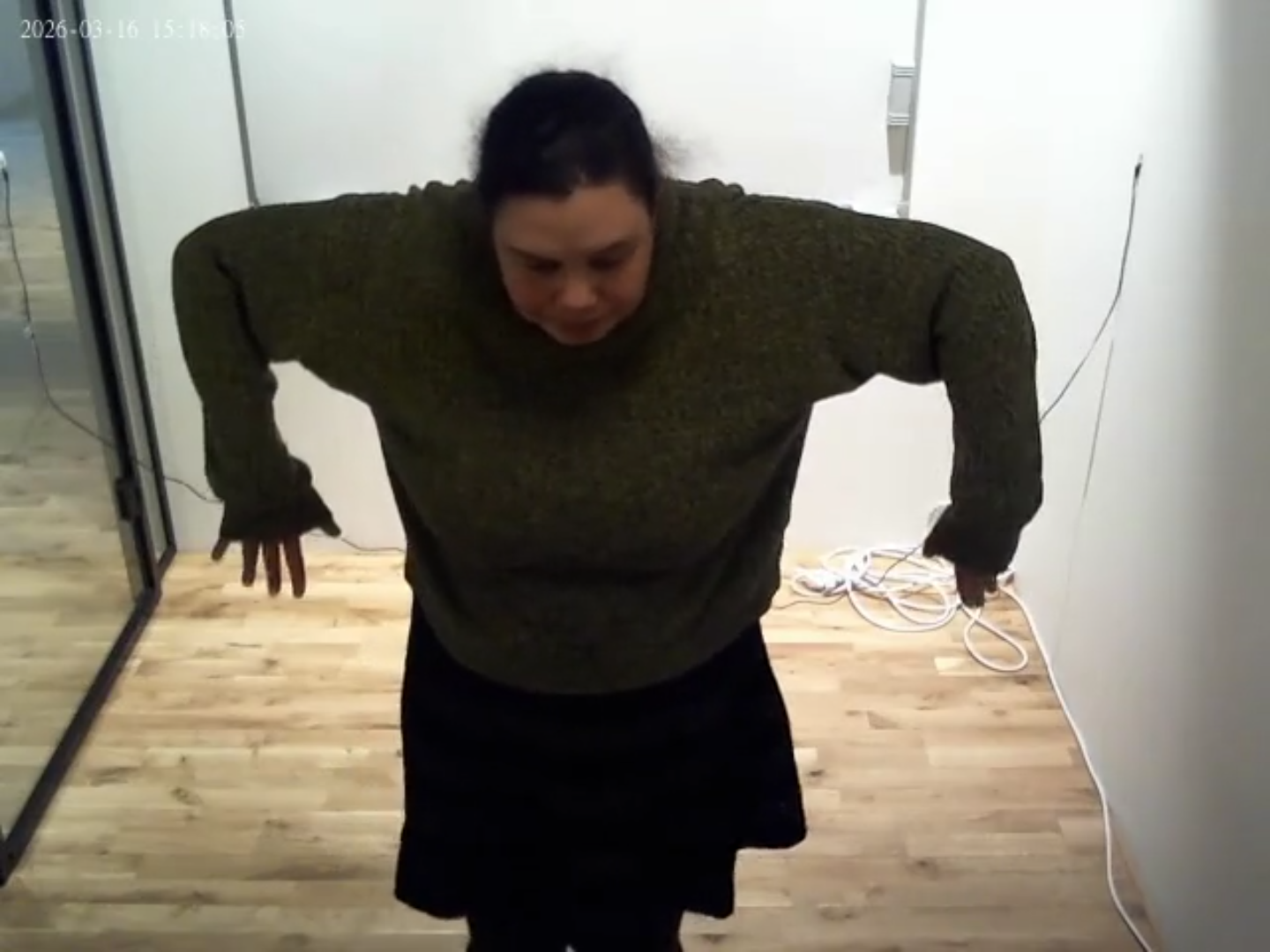}
        \caption*{\small \centering Rotating with \\
        bent arms}
    \end{minipage}
    \begin{minipage}{0.32\columnwidth}
        \includegraphics[width=\linewidth]{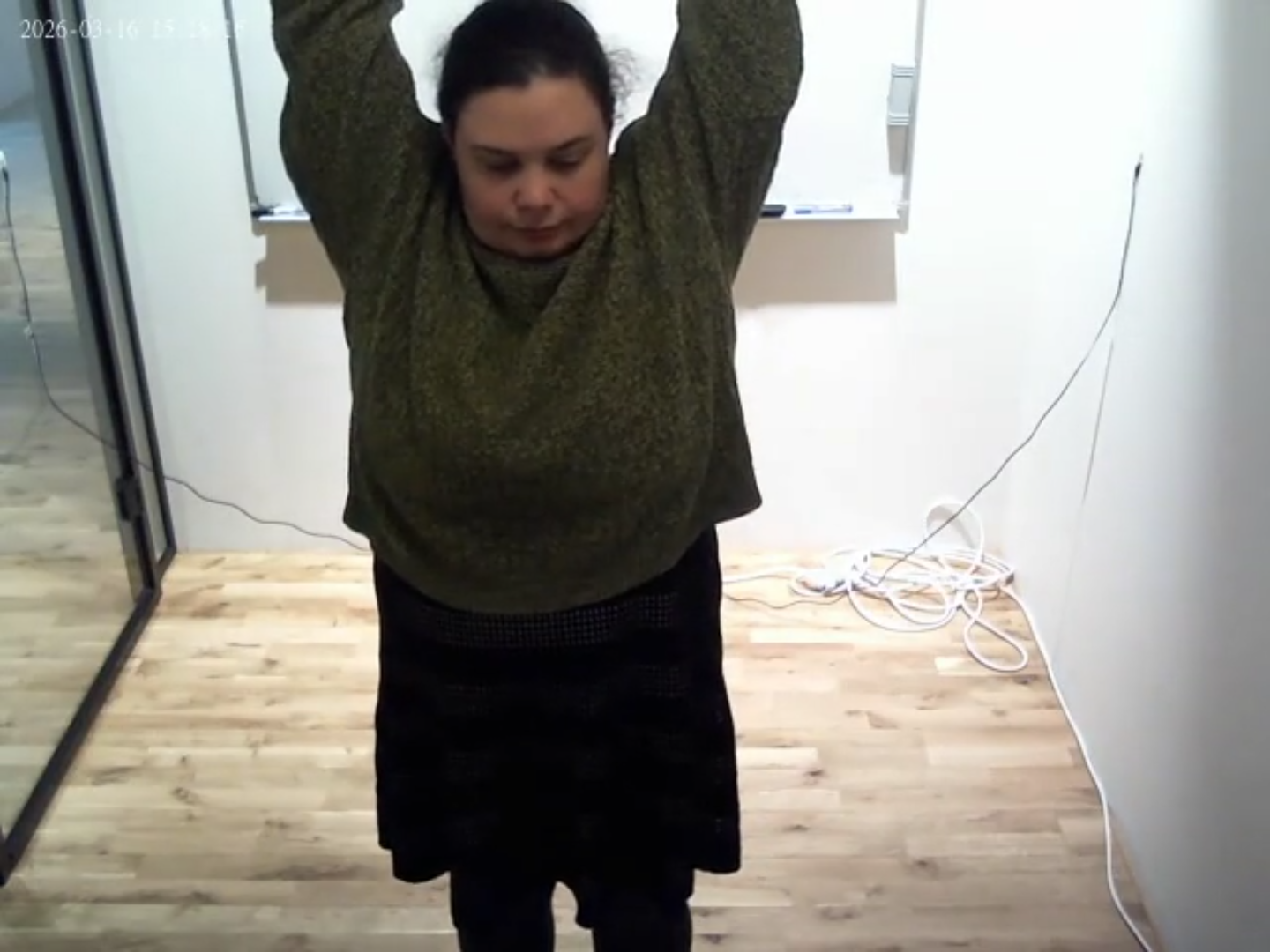}
        \caption*{\small \centering Raising \\
        both arms}
    \end{minipage}\vspace*{-0.75ex}
    \caption{Example images from video of each event in Table~\ref{tab:annotations}}
    \label{fig:video_label_stills}
\end{figure}

In this section we present the results and evaluation of our cross-domain approach presented in Section~\ref{sec:Crossdomain}. To correlate the RSSI data with the captured video (i.e.\ our ground truth), Table~\ref{tab:annotations} lists the human movement events, where some example video frames are given in \figurename~\ref{fig:video_label_stills}.

\subsection{Event Ordering}
\begin{figure*}[!t]
    \centering
    \includegraphics[width=1.0\linewidth, trim=0pt 0pt 0pt 0.75cm, clip]{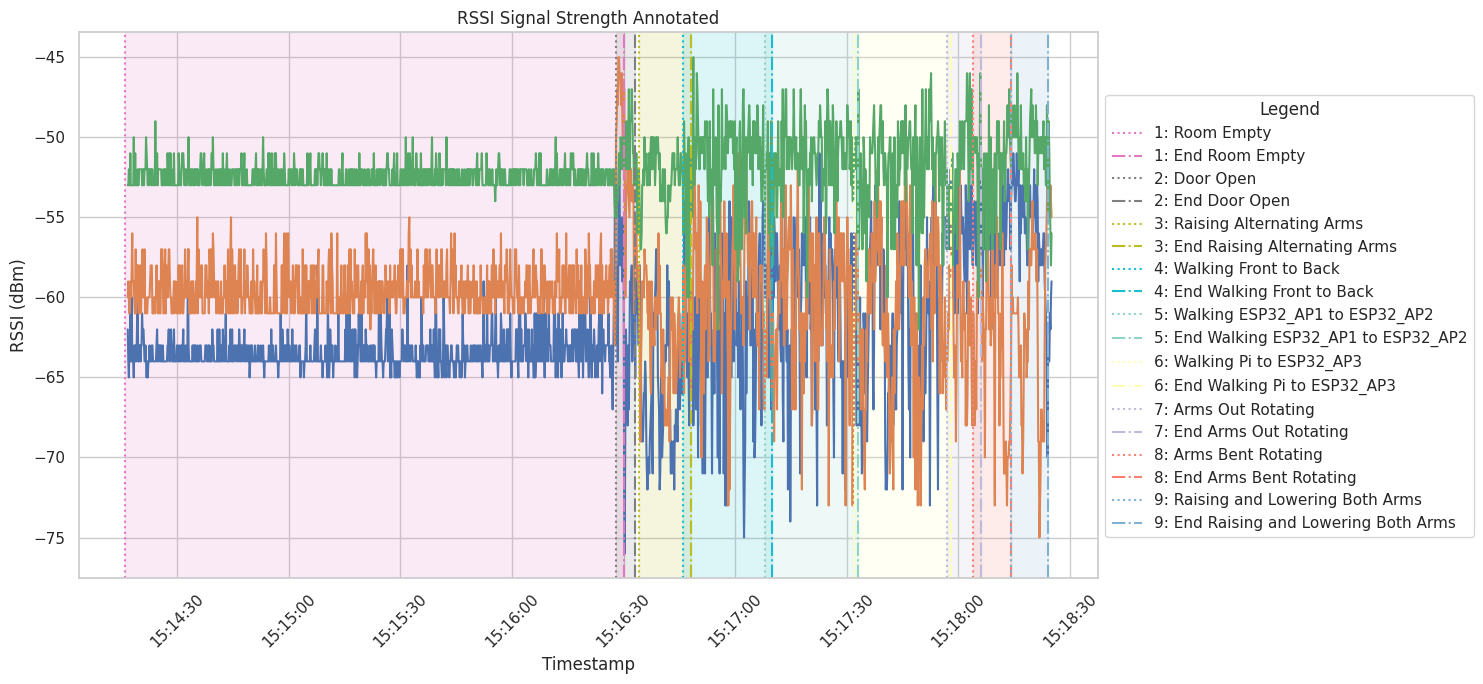}\vspace*{-1.5ex}
    \caption{RSSI signal strength with the time range of actions labeled.}
    \label{fig:annotated_data}
\end{figure*}

Using the timestamps from Table~\ref{tab:annotations}, the \ac{RSSI} data was annotated, as shown in Fig.~\ref{fig:annotated_data}. %
The colors and labels assigned to each event in Table~\ref{tab:annotations} and Fig.~\ref{fig:annotated_data} are used consistently throughout the remaining figures. 

\begin{table}[!b]
    \centering
    \caption{Event Timestamps}
    \label{tab:annotations}
    \begin{tabular}{ll}
        \toprule
        Event & Time Stamp \\
        \midrule
        1: Room Empty & 15:14:16.696-15:16:29 \\
        2: Door Open & 15:16:29-15:16:32 \\
        3: Raising Alternating Arms & 15:16:35-15:16:47 \\
        4: Walking Front to Back & 15:16:47-15:17:09 \\
        5: Walking ESP32\_AP1 to ESP32\_AP2 & 15:17:09-15:17:32 \\
        6: Walking Pi to ESP32\_AP3 & 15:17:33-15:17:57 \\
        7: Arms Out Turning & 15:17:58-15:18:05 \\
        8: Arms Bent Turning & 15:18:05-15:18:13 \\
        9: Raising/Lowering Both Arms & 15:18:15-15:18:23 \\
        \toprule
    \end{tabular}
\end{table}

\begin{table}[!b]
\centering\vspace*{-1ex}
\caption{Skeletal Keypoint Connections and Body Regions}
\label{tab:keypoints}
\scriptsize
\begin{tabular}{lll}
\toprule
\textbf{Region} & \textbf{Connection $(A, B)$} & \textbf{Description} \\
\midrule
\textbf{Head} & (12, 13) & Head to Neck \\
\midrule
\multirow{3}{*}{\textbf{Torso}} & (10, 6) & Upper Torso to Left Hip \\
& (6, 8) & Left Hip to Right Hip \\
& (8, 11) & Right Hip to Upper Torso \\
\midrule
\multirow{6}{*}{\textbf{Arms}} & (13, 10) & Neck to Left Shoulder \\
& (10, 7) & Left Shoulder to Left Elbow \\
& (7, 3) & Left Elbow to Left Wrist \\
& (13, 11) & Neck to Right Shoulder \\
& (11, 9) & Right Shoulder to Right Elbow \\
& (9, 5) & Right Elbow to Right Wrist \\
\midrule
\multirow{4}{*}{\textbf{Legs}} & (6, 2) & Left Hip to Left Knee \\
& (2, 0) & Left Knee to Left Ankle \\
& (8, 4) & Right Hip to Right Knee \\
& (4, 1) & Right Knee to Right Ankle \\
\bottomrule
\end{tabular}
\end{table}

\begin{figure*}[!t]
    \centering\vspace*{-3.4ex}
    \includegraphics[width=1.0\linewidth]{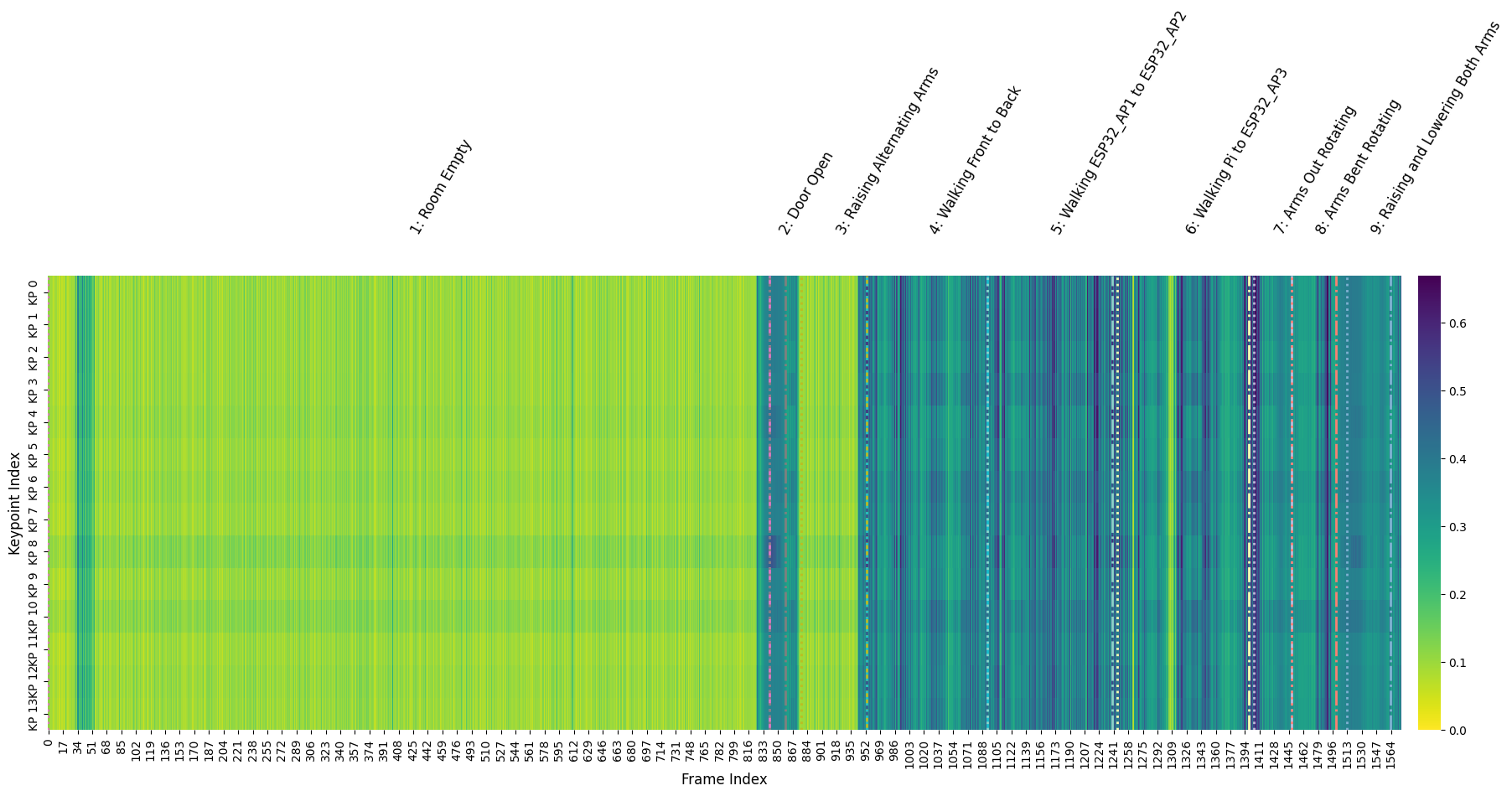}\vspace*{-1.5ex}
    \caption{Standard deviation between each frame's 100 predictions.}\vspace*{-3.5ex}
    \label{fig:keypoint_std}
\end{figure*}

\subsection{Prediction Correlation Analysis}

As the \emph{Person-in-WiFi 3D} model makes 100 predictions per frame for each skeletal keypoint we perform statistical analysis on each keypoint.
The standard deviation of each keypoint, representing a joint in the pose model, remains in the same standard deviation range when the room is empty, as depicted in \figurename~\ref{fig:keypoint_std}. The lower the standard deviation value, the more closely the keypoints from all the 100 predictions are in each frame. A lower value indicates a higher level of consensus between the location of the predictions. By combining this with the confidence scores, we can determine the predictions were precise based on the model's understanding of \ac{CSI} data, even if the location is inaccurate compared to reality. 

Figure~\ref{fig:keypoint_std} in combination with Table~\ref{tab:keypoints} shows that the keypoints with the highest standard deviation throughout the predictions are 6, 8, and 10 (i.e.\ the visible horizontal bars in \figurename~\ref{fig:keypoint_std}); all points are part of the torso region, i.e.\ where the human body has most of its mass that influences  Wi-Fi signals, e.g.\ via Doppler effect. There is also an observable vertical bar of low standard deviation frames when the door was opened to allow the person into the lab space. We would expect the model to be confused by the change in distance between the ESP32\_AP3 transmitter and the receiver raising the signal strength. It may also perceive the change as evidence an object has entered the space and changed the reflectance of broadcast packets.

To try and improve the delineation between these two states, we applied a minimum threshold and sigmoid value to remove noise and smooth the data in preparation for scaling to mimic the \ac{CSI} \emph{phase denoising} performed in the \emph{Person-in-WiFi 3D} project. The angular embedding was then applied by subtracting 0.5 to center the sigmoid value and the multiplying it by $\pi^2$, 
see Listing~\ref{eq:sigmoid}. The embedding is based on a unit circle with a point traveling around its perimeter, a distance of less than $\pi$ from the circle's center indicates the signal strength was lower than expected. A noise threshold of 0.00095 and K value of 4000 were selected to preserve the model's responsiveness to fluctuations while quieting the signal. After this, values will fall within $\pm\pi$ as expected for phase data and represents the voltage observed at the antenna.

\begin{figure}
    \centering
    \includegraphics[width=1.0\linewidth]{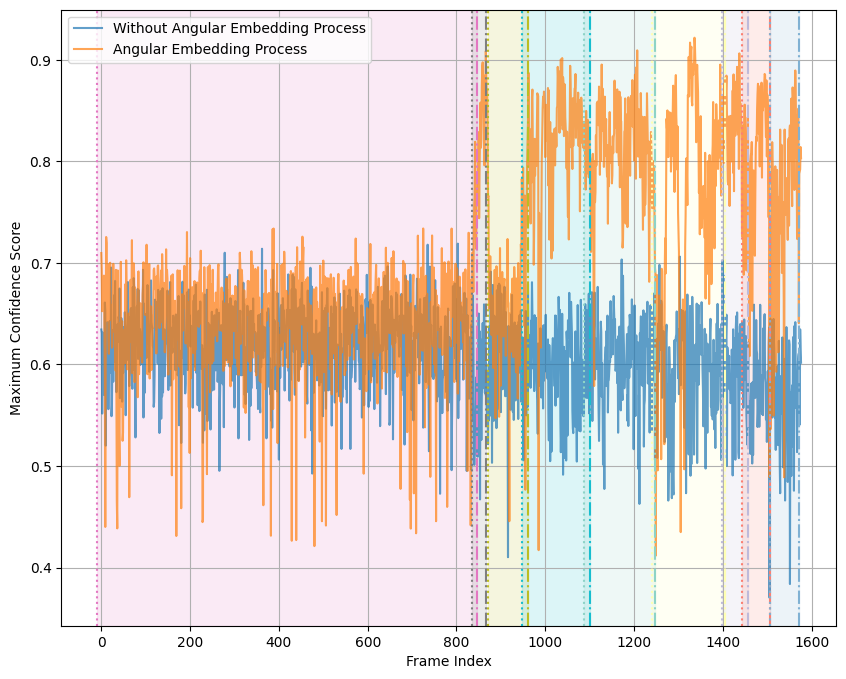}\vspace*{-0.6ex}
    \caption{Change in maximum confidence for each frame when a noise gate, smoothing, and angular embedding are performed.}
    \label{fig:max_angular}
\end{figure}

\begin{figure}
    \centering
    \includegraphics[width=1.0\linewidth]{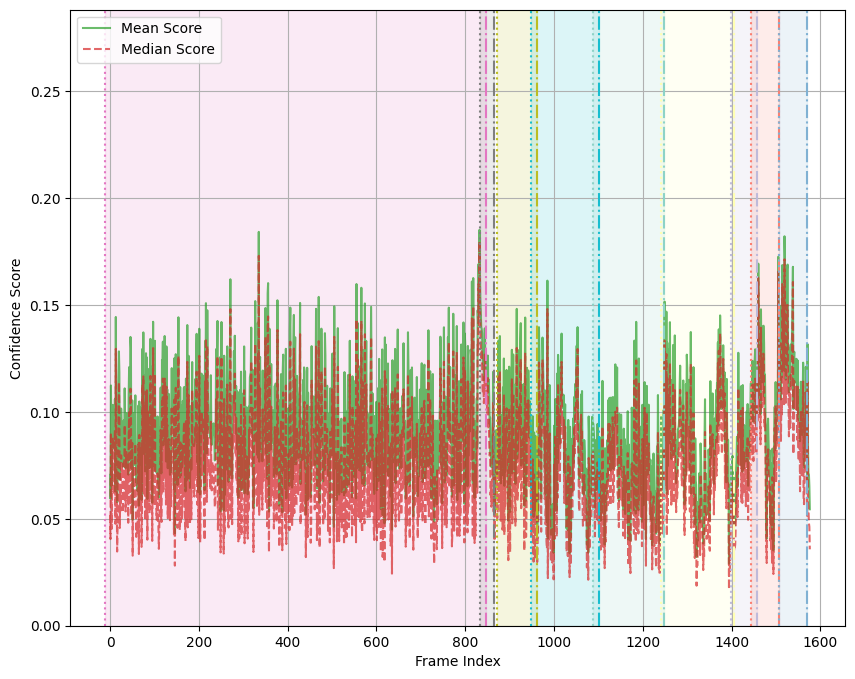}
    \caption{Mean and median of the results dataset with angular embedding.}\vspace*{-1ex}
    \label{fig:mean_avg_angular}
\end{figure}

\begin{figure}
    \centering
    \includegraphics[width=1.0\linewidth]{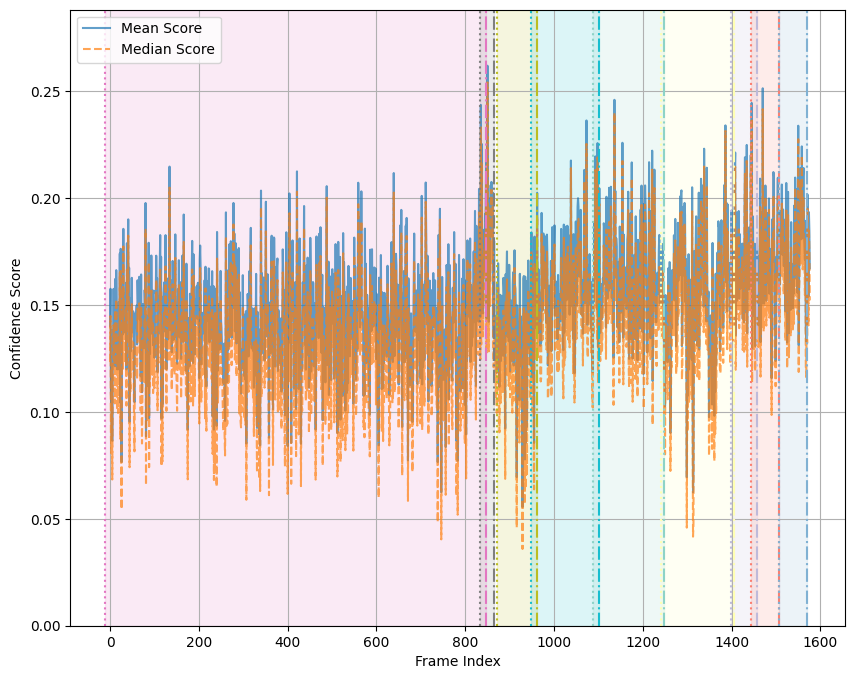}
    \caption{Mean and median of the results dataset with  subtraction of global mean only.}
    \label{fig:mean_avg_feasibility}
\end{figure}

\begin{lstlisting}[language=Python, float=t, belowskip=-2ex,
caption={Angular embedding Python code}, label={eq:sigmoid}]
filtered = np.abs(diff) - noise_threshold
diff_centered = np.sign(diff) * np.maximum(0,filtered)
sig = 1 / (1 + np.exp(-K * diff_centered))
return (np.pi ** 2) * (sig - 0.5)
\end{lstlisting}

\figurename~\ref{fig:max_angular} plots the maximum confidence score for pre-processed data with and without angular embedding. The figure shows that the confidence of the predictions in the empty room are nearly the same as for the occupied room. Comparing the two, we can see that the maximum confidence values for the empty and occupied room are clearly separated into ranges with limited overlap. The mean and median of the predictions, \figurename~\ref{fig:mean_avg_angular} and~\ref{fig:mean_avg_feasibility}, are also lower on average with the increased amplitude of the angular embedded data, indicating the prediction confidence scores are based on a feature of the data, not of the amplitude of the input. This suggests that the noise of the \ac{RSSI} data has incorrectly been perceived as an object moving in the collection space. Therefore, the angular embedding had the desired effect of increasing the occupied room's prediction confidence, as shown by the orange line in Fig.~\ref{fig:max_angular}. However, the mean and median scores were lowered as a result as seen in \figurename~\ref{fig:mean_avg_angular}. This is in contrast to the predictions based on the pre-processing method shown as the blue maximum confidence score line in \figurename~\ref{fig:max_angular} and the mean and median graph of \figurename~\ref{fig:mean_avg_feasibility}. This is to be expected as the original training method was a bipartite matching loss detection transformer. The normal training pipeline creates the 100 raw predictions exactly as our implementation does, but the prediction that matches the ground truth is given a higher score by the teaching model. The student pose decoder model produces a higher score for predictions it perceives as matching the pose being made in the collection space as a result of this reinforcement.

Overall, the maximum prediction confidence score reached 0.71894246, as depicted by the blue maximum confidence score line in \figurename~\ref{fig:max_angular}. This is higher than random guessing and demonstrates the model's evaluation process found the \ac{RSSI} signal data to be generally as meaningful as \ac{CSI}. With the use of angular embedding, the maximum prediction confidence score, as shown by the orange maximum confidence score line in \figurename~\ref{fig:max_angular}, reached 0.92170554 with roughly 0.8, or 80\%, average confidence when the room is occupied.

Therefore, the correlation of labeled events with the RSSI and predictions (Table~\ref{tab:annotations} and Fig.~\ref{fig:annotated_data}), indicate the possible use of \ac{RSSI} data for predicting the location of novel objects and people. This is further supported by the high confidence scores (\figurename~\ref{fig:max_angular}), and the decrease in standard deviation (\figurename~\ref{fig:keypoint_std}) when the room is occupied. Together this supports predication via a model from an existing Wi-Fi positioning or pose prediction project without additional training.

\section{Discussion}\label{sec:Discussion}

In combination, the results have shown that the statistical behavior of the raw predictions supports the possible use of \ac{RSSI} data for predicting human presence and movement.
\\\textbf{Implications for Privacy}.
The results of this research suggest individual privacy is at risk of possible location tracking in \ac{IoT}-rich environments using consumer hardware for \ac{RSSI} data collection. This is in contrast to \ac{CSI} data collection methods that require lower-level permissions in the operating system to access. Related works focus on \ac{CSI} data makes the use of consumer Wi-Fi adapters less feasible as many do not expose \ac{PHY}-layer information. While legacy drivers in Linux can sometimes expose such functionality, manufacturers do not provide access in consumer equivalent drivers. As an alternative to the single-core ESP32-C3, the use of ESP32 Xtensa devices would be an accessible alternative to specialized Wi-Fi adapters for \ac{CSI} collection. These devices are representative of a wide range of IoT devices on the market and provide multi-core processors with higher processing speeds.
\\\textbf{Threats to Validity.}
Internal threats to validity focus on the physical experimental setup. The space used in the \emph{Person-in-WiFi 3D} lab environment was set up in an open floor-plan space; this is in strong contrast to the small concrete room used for this project's lab. The open area reduces general reflectance of the space and leads to less background noise. It is possible the echo effect of the Wi-Fi signal bouncing off of objects and walls repeatedly makes predictions less accurate and the lack of granular information in \ac{RSSI} data compounds this. We accounted for this by subtracting away the noise from the empty room. 
External threats to validity include the lack of alignment with the \emph{Person-in-WiFi 3D} model's focus on \ac{CSI} data during training in combination with the unavailability
of a Kinect Azure or equivalent device. To address the limitation of access to a Kinect Azure for ground truth generation, a secondhand version of the Kinect could be used and the depth or field for the infrared vision could be supplemented with an additional LED array and sensor(s). This would likely further increase the need for compute power during collection as the Kinect Azure development kit appears to synchronize its sensor data before forwarding the captured content. 

\section{Conclusion} \label{sec:Conclusion}
With the increased use of Wi-Fi and other wireless communications in everyday devices, such as \ac{IoT} devices, the ambient surveillance vectors grow as well. Traditional methods of surveillance rely on cameras and microphones, features that are well understood and controlled by the average user. In contrast, wireless signals are ephemeral and invisible, making them far less apparent as potential tools for monitoring. This dichotomy of ubiquity and ambiance makes the design and auditing of smart device \ac{SDK}s a public cybersecurity obligation.

By investigating the feasibility of using \ac{RSSI} data as input for the \ac{CSI}-focused \emph{Person-in-WiFi 3D} model, this research has quantified the confidence and precision that ambient signal data can elicit from a model trained with \ac{CSI} data. This demonstrates low-density signal data can still trigger high-certainty predictions with a low degree of deviation, proving the obscurity of the vector does not protect the user, but instead necessitates a focus on secure design that prioritizes privacy.

\section*{Acknowledgment}
We would like to thank the authors of~\cite{10656837} for sharing their model and training data with us.
This project has received co-funding from the Icelandic government as well as from the European Union's Digital Europe Programme and the European Cybersecurity Competence Centre under grant agreement no.\ 1011226821 Eyvör National Coordination Centre for Cybersecurity Iceland and 101127307 Defend Iceland: Nationwide bug bounty platform.

\bibliographystyle{IEEEtran}
\bibliography{references}

\end{document}